\documentclass[aos,MSNbibl,seceqn,dvips]{arximspdf}
\usepackage{multirow}
\usepackage{graphicx}

\doi{10.1214/12-AOS1076} 
\volume{41}
\issue{1}
\pubyear{2013}
\firstpage{125}
\lastpage{142}

\makeatletter
\renewcommand{\citep}[1]{(\citeauthor{#1} \citeyear{#1})}
\newtheorem{thmm}{Theorem}[section]
\newtheorem{lem}[thmm]{Lemma}
\newtheorem{algorithm}{Algorithm}

\def\Joint{\mathcal{J}}
\def\Z{\mathbb{Z}}
\def\Opt{\mathcal{O}}
\def\Blind{\mathcal{B}}
\def\Unr{\mathcal{U}}
\def\Int{\mathcal{I}}
\def\Pilot{\mathcal{P}}
\def\mG{\mathcal{G}}
\def\mC{\mathcal{C}}
\def\rd{\mathrm{d}}
\def\R{\mathbb{R}}
\def\E{\mathrm{E}}
\def\N{\mathbb{N}}
\def\Z{\mathbb{Z}}
\def\Prob{\mathrm{P}}
\def\Ind{\mathbb{I}}
\makeatother

\begin{document}
\begin{frontmatter}

\title{An algorithm to compute the power of Monte Carlo tests with
guaranteed precision\thanksref{T1}}
\runtitle{Computing the power of Monte Carlo tests}

\begin{aug}
\author[A]{\fnms{Axel} \snm{Gandy}\corref{}\ead[label=e1]{a.gandy@imperial.ac.uk}}
\and
\author[B]{\fnms{Patrick} \snm{Rubin-Delanchy}\ead[label=e2]{patrick.rubin-delanchy@bristol.ac.uk}}
\thankstext{T1}{Supported by EPSRC Grant EP/H040102/1.}
\runauthor{A. Gandy and P. Rubin-Delanchy}
\affiliation{Imperial College London and University of Bristol}
\address[A]{Department of Mathematics\\
Imperial College London\\
South Kensington Campus\\
London SW7 2AZ\\
United Kingdom\\
\printead{e1}}
\address[B]{School of Mathematics\\
University of Bristol\\
University Walk\\
Bristol BS8 1TW\\
United Kingdom\\
\printead{e2}}
\end{aug}

\received{\smonth{3} \syear{2012}}
\revised{\smonth{8} \syear{2012}}


\begin{abstract}This article presents an algorithm that generates a
conservative confidence interval of a specified length and coverage
probability for the power of a Monte Carlo test (such as a bootstrap
or permutation test). It is the first method that achieves this aim
for almost any Monte Carlo test. Previous research has focused on
obtaining as accurate a result as possible for a fixed computational
effort, without providing a guaranteed precision in the above sense.
The algorithm we propose does not have a fixed effort and runs until a
confidence
interval with a user-specified length and coverage probability can
be constructed. We show that the expected effort required by the
algorithm is finite in
most cases of practical interest, including situations where the
distribution of the $p$-value is absolutely continuous or discrete
with finite support. The algorithm is implemented in the
R-package \textit{simctest}, available on CRAN.
\end{abstract}

%

\end{frontmatter}
%
\section{Introduction}\label{sec1}
Let $p$ be a random variable taking values in $[0,1]$ with unknown
cumulative distribution function (CDF) $F$. For some $\alpha\in(0,1)$,
we want to approximate $\beta=F(\alpha)$ by Monte Carlo
simulation. Assume that we cannot sample from $F$ directly, but that
it is possible to generate a collection of random variables $(X^i_j\dvtx i
\in\N, j \in\N)$, where $X^i_1,X^i_2, \ldots\sim$ Bernoulli$(p_i)$
independently
and $p_1,p_2,\ldots$ are unobserved independent copies of $p$,
that is, $p_{1}, p_2,\ldots\sim F$ independently.

This problem comes about when computing the power or level of a Monte
Carlo test, such as a bootstrap or permutation test, or in general a
test that rejects on the basis of simulations under the (potentially
estimated) null hypothesis. In this context, $p$ is the (random)
$p$-value, $\alpha$ the nominal level of the test and $\beta$ its
power.
In this situation $X^i_1, X^i_2,\ldots$ are generated as follows:
simulate a dataset (thus implicitly generating $p_i$), compute the
observed test statistic and then, for $j=1,2,\ldots,$ use a sampling
technique (such as bootstrapping or permutation) on the observed
dataset to get a (re)sampled realization of the test statistic under
the null hypothesis. Define $X^i_j$ as the indicator that the
(re)sampled test statistic is at least as extreme as the observed test
statistic.

A typical approach
is to choose $N,M\in
\N=\{1,2,\ldots\}$ and estimate $\beta$ by
\[
\hat{\beta}_{\mathrm{na\ddot{\i}ve}} = \frac{1}{N}\sum_{i=1}^N
\Ind \Biggl[ \Biggl(\frac{1}{M}\sum_{j=1}^M
X^i_{j} \Biggr) \leq\alpha \Biggr],
\]
where $\Ind$ is the indicator function. A problem of this approach
is that the bias of $\hat{\beta}_{\mathrm{na\ddot{\i}ve}}$ is unknown. For
example, using \cite{boos00}, equation (2), it can be shown that no matter
how large $N$ and $M$ are chosen,
\[
\mathop{\sup}_{\Prob\in\mathcal{P}}|\E\hat{\beta}_{\mathrm{na\ddot{\i}ve}}-\beta| \geq0.5,
\]
where $\mathcal{P}$ is the set of all probability distributions on
$[0,1]$. Better bounds are available under the assumption that $\E\hat
{\beta}_{\mathrm{na\ddot{\i}ve}}$ is concave in $\alpha$, see \cite{davison1997bma},
Section~4.2.5. However, this would usually not be known in a
given application.

More advanced estimation methods have been proposed. For instance, Oden~\cite{Oden91}
has investigated how to choose the relative sizes of $N$
(controlling the variance) and $M$ (controlling the bias), to minimize
the total estimation error for certain distributions of $p$. \cite
{boos00} partially correct the bias by extrapolation.

However, existing procedures do not provide a formal, finite-sample
guarantee on the accuracy of $\hat{\beta}$ for a general test. This is
partly because the problem has always been approached with the
principle of finding as accurate an estimate as possible for a fixed
computational effort.

We approach the problem with the priorities reversed: we make an exact
probabilistic statement about the result, allowing the computational
effort to be random.

The algorithm that we propose is guaranteed to provide a conservative
confidence interval (CI) for $\beta$ of a given coverage probability.
This interval will, after a finite expected number of samples, reach
any desired length, provided that $F$ is H\"older continuous in a
neighborhood of $\alpha$ with exponent $\xi> 0$. This is satisfied if,
for example, in a neighborhood of $\alpha$, $p$ is absolutely
continuous with respect to Lebesgue measure with bounded density. In
this case $\xi=1$.

For practical use, the inner workings of the algorithm can be
ignored. Users only need to provide the required precision (maximum
CI length and minimum coverage probability) and a
mechanism for generating the $X^i_j$. The algorithm is implemented in
the R-package \textit{simctest}, available on CRAN.

The article is structured as follows. In Section~\ref{secmain} we
describe the basic algorithm. Theorem
\ref{thmfinstoptime} demonstrates that, under very mild conditions,
the algorithm
terminates in finite expected time. Sections~\ref{secnostreams} and
\ref{secseqhyp} present additional methodology to reduce the
computational effort, some details of which are in supplementary material~\cite
{supplement}. Section~\ref{secsimulations} contains a simulation
study. In Section~\ref{secadaptci}, we suggest an adaptive rule which
ensures that the computational effort is only high if the estimate is
in a region of interest. In Section~\ref{secpermtest} we demonstrate
the use of our algorithm on a simple permutation test example. Proofs
and auxiliary lemmas are in the \hyperref[app]{Appendix}. Within these, Lemma~\ref
{lem1} confirms an observation made in \cite{gandy09}, main text page~1507 and
Figure 4, about the distance between certain stopping boundaries.

\section{The basic algorithm}\label{secmain}
\subsection{Description}
We use the notation introduced in the first paragraph of the
\hyperref[sec1]{Introduction}.
For every $i\in\N$, we call the Bernoulli sequence $(X^i_j)_{j\in\N}$
a
\emph{stream}.
The algorithm will use a fixed number $N$ of these
streams.

For each stream $i$, our algorithm aims
to decide if $p_i\leq\alpha$ or if $p_i>\alpha$. We use the sequential
algorithm of \cite{gandy09} for this purpose.

To simplify notation, we often drop the stream index $i$ when referring
to a generic stream; for example, we write $X_j$, $p$ instead of
$X_j^i$, $p_i$. Furthermore, we
use a subscript to indicate the probability distribution of such a
stream conditional on a specific value of $p$, that is, $\Prob_q(\cdot
)=\Prob(\cdot|p=q)$ for some $q\in[0,1]$.

The procedure in \cite{gandy09} defines two deterministic sequences, an
upper boundary $(U_t\dvtx t \in\N)$ and a lower boundary $(L_t\dvtx t \in\N
)$. While the partial sum $S_t = \sum_{j = 1}^t X_j$ has hit
neither boundary, the stream is \emph{unresolved}. The procedure
terminates at the hitting time
\[
\tau= \inf\{t\dvtx S_t \geq U_t \mbox{ or }
S_t \leq L_t\}.
\]
If the upper boundary is hit, we decide $p>\alpha$ and report a \emph
{negative outcome} ($p$ is not significant at level $\alpha$). If the
lower boundary is hit we decide $p\leq\alpha$ and report a \emph
{positive outcome} ($p$ is significant at level $\alpha$).

\begin{figure}

\includegraphics{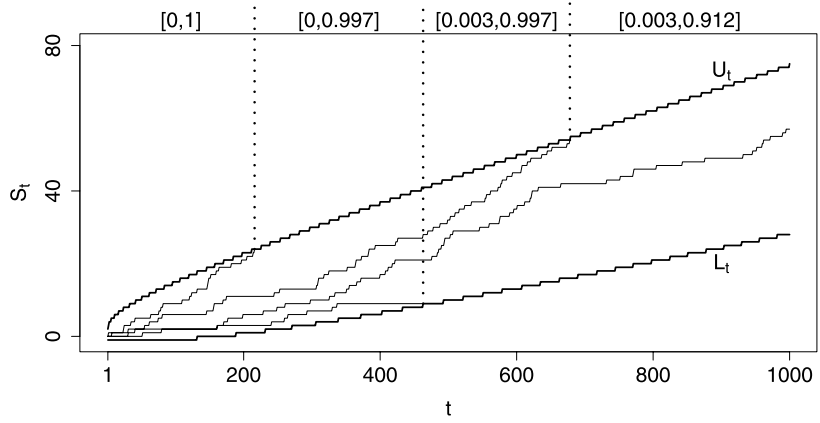}

\caption{Confidence intervals generated by the algorithm using $N=4$,
$\varepsilon=0.01$, $\alpha=0.05$, $\varepsilon_t= \varepsilon t/(1000+ t)$ and
$\gamma=0.05$.}\label{figIllustralg}\vspace*{-3pt}
\end{figure}

The boundaries are constructed to give a desired uniform bound
$\varepsilon
>0$ on the probability of a wrong decision, that is,
%
\begin{eqnarray}
\label{eqhitprob} %
\Prob_p(S_{\tau}=U_{\tau})&
\leq&\varepsilon\qquad\mbox{for } p\leq\alpha,
\nonumber
\\[-8pt]
\\[-8pt]
\nonumber
\Prob_p(S_{\tau}=L_{\tau})&\leq&\varepsilon\qquad\mbox{for } p> \alpha.
\end{eqnarray}
Figure~\ref{figIllustralg} shows an example of $U_t$ and $L_t$ with
$\varepsilon= 0.01$
and $\alpha=0.05$.

To be more precise, the boundaries are constructed recursively using a
\emph{spending sequence}
$(\varepsilon_t)$ with $0\leq\varepsilon_t \nearrow
\varepsilon$ as $t\rightarrow\infty$. The spending sequence governs how
quickly the error probability $\varepsilon$ is spent, guaranteeing
\begin{eqnarray*}
\Prob_p(S_{\tau}=U_{\tau}, \tau\leq t)&\leq&
\varepsilon_t\qquad \mbox{for } p\leq\alpha,
\\
\Prob_p(S_{\tau}=L_{\tau}, \tau\leq t )&\leq&
\varepsilon_t\qquad \mbox{for } p> \alpha.
\end{eqnarray*}
The precise recursive construction is given in (\ref{equldef}), in the
\hyperref[app]{Appendix}.\vadjust{\goodbreak}

Our algorithm runs $N$ streams in parallel until enough have been
resolved to meet the required precision. More formally, it operates as follows:

\begin{algorithm}[(Basic algorithm)]
\label{algbasic}
\begin{tabbing}
Let $t=0$; $R_0=0$; $A_0=0$; $\Unr_0=\{1,\ldots,N\}$, $S^1_0=0,\ldots
,S^N_0=0$ \\
while \= $|I(R_t, A_t, |\Unr_t|;\gamma)| > \Delta$\\
\>Let $t=t+1$, $R_t = R_{t-1}$, $A_t=A_{t-1}$, $\Unr_t =
\Unr_{t-1}$ \\
\>for \=$i\in\Unr_t$ \\
\>\>Let $S^{i}_t=S^{i}_{t-1}+X^i_{t}$\\
\>\>If $S^{i}_t\geq U_t$ let $A_t=A_t+1$, $\Unr_t= \Unr_{t}\setminus
\{
i\}$\\
\>\>If $S^{i}_t\leq L_t$ let $R_t= R_{t}+1$, $\Unr_{t}= \Unr_{t}\setminus\{i\}$\\
Report $I(R_t, A_t, |\Unr_t|;\gamma)$ as confidence interval for
$\beta$.
\end{tabbing}
\end{algorithm}
$\Unr_t$ is a set containing the indices of unresolved streams at
time $t$. $|\cdot|$ denotes the size of finite sets as well as the
length of intervals. $R_t$ and $A_t$ count, respectively, the number of
positive and negative outcomes.

$I(R_t, A_t, |\Unr_t|;\gamma)$ is a conservative confidence interval
for $\beta$ based on $R_t$, $A_t$ and $|\Unr_t|$. It is constructed as
follows. Because of (\ref{eqhitprob}), the probability that a stream
has a positive outcome is in the interval $[(1-\varepsilon)\beta,
(1-\varepsilon)\beta+ \varepsilon]$. Therefore, if all streams were
resolved, the following interval would be a conservative confidence
interval for $\beta$ with coverage
probability $1-\gamma$:
\[
\Int_{\infty}=\Int_{\infty}(R_{\infty},A_{\infty};
\gamma)= \biggl[ \frac
{\beta_-^* - \varepsilon}{1-\varepsilon}, \frac{\beta^*_+}{1-\varepsilon
} \biggr],
\]
where $R_{\infty}$ ($A_{\infty}$) denotes the number of positive
(negative) outcomes and $[\beta_{-}^{*}, \beta_{+}^{*}]$ is the
Clopper--Pearson confidence interval \cite{clopper34} with coverage
probability $1-\gamma$ for the success probability of a Binomial random
variable observed to be $R_{\infty}$ after $R_{\infty} +A_{\infty}$ trials.\vadjust{\goodbreak}

The subscript in $\Int_{\infty}$ represents that this is the interval
that would be obtained by our algorithm if it were run
until all streams were resolved.

To obtain a conservative confidence interval $\Int_t$ while there are
unresolved streams, we take the union of all confidence intervals that
could be obtained after observing the outcomes of the remaining
streams, that is, we let $\Int_t = I(R_t, A_t, |\Unr_t|; \gamma)$ where
%
\begin{equation}
\label{defI} I(r, a, u; \gamma) = \bigcup_{r_{\infty} = r}^{r + u}
\Int_{\infty
}(r_{\infty}, r + a + u - r_{\infty}; \gamma).
\end{equation}
By construction, $\Int_1\supseteq\Int_2\supseteq\cdots\supseteq
\Int_{\infty}$ and
\[
\Prob[\beta\in\Int_1 \cap\cdots\cap\beta\in\Int_{t}\cap
\cdots\cap \beta\in\Int_{\infty}] \geq1-\gamma.
\]

Figure~\ref{figIllustralg} illustrates the algorithm in a toy example
with only $N=4$ streams. The thin lines depict the 4 corresponding
partial sum sequences, $S^{i}_t$. When $S_t^{i}$ hits one of the
boundaries the stream is stopped, causing a retraction of the
confidence interval for $\beta$ (annotated at the top of the graph).

\subsection{Expected time}
A simpler algorithm than Algorithm~\ref{algbasic} would be to run $N$
streams until all are resolved. $N$ can be chosen such that the CI
length is at most~$\Delta$ for all outcomes. However, this algorithm is
unusable in practice as it typically requires an infinite expected
effort. Indeed, from \cite{gandy09}, page 1506, if the CDF $F$ of~$p$
has a nonzero derivative at $\alpha$, a very common case, then
$\E[\tau_{i}] = \infty$, where $\tau_i$ denotes the hitting time of the
$i$th stream. This makes the overall expected effort infinite.

We now show that with our algorithm we can choose $N$ and $(\varepsilon_{t})$ such that the expected effort is finite. The key is to make $N$
large enough that not all streams have to be resolved.

The effort of Algorithm~\ref{algbasic}, as measured by the number of
$X^i_t$ used, is
%
\begin{equation}
e=\sum_{i=1}^{N} \min\{
\tau_{i}, \tau_{(N-k)}\}, \label{eqeffort}
\end{equation}
where $k$ is the number of unresolved streams when the algorithm
finishes and $\tau_{(1)} \leq\cdots\leq\tau_{(N)}$ denote the order
statistics of $\tau_1, \ldots, \tau_N$.

By choosing $N$ large enough and $\varepsilon$
small enough, we can ensure $k\geq\kappa$ for any given $\kappa\geq1$.
The effort is then bounded above by $\tau_{(N-\kappa)} N$. Thus to
ensure that $\E[e]$ is finite, it suffices to prove $\E[\tau_{(N-\kappa
)}] < \infty$ for some $\kappa$. The following theorem shows that in
many cases $\kappa$ can be taken as small as $2$.

\begin{thmm}\label{thmfinstoptime}
Suppose that $\varepsilon\leq1/4$ and there exist constants $\lambda>
0$, $q>1$ and $T \in\N$ such that $\varepsilon_t - \varepsilon_{t-1} \geq
\lambda t^{-q}$ for all $t \geq T$. Further, suppose that in a
neighborhood of $\alpha$ the CDF $F$ of $p$ is H\"older continuous with
exponent $\xi$. Then $\E[\tau_{(i)}] < \infty$ for $i \leq N -
\lfloor
2/\xi\rfloor$. In particular, if $\xi= 1$ (the CDF is Lipschitz
continuous in a neighborhood of $\alpha$), then $\E[\tau_{(N-2)}] <
\infty$.
\end{thmm}

$F$ is H\"older continuous with exponent $\xi$ in a neighborhood of
$\alpha$ if there exists an open interval $U$ containing $\alpha$ for
which there exists a $c>0$ such that for all $x,y\in U$, $|F(x)-F(y)|
\leq c |x-y|^{\xi}$.

The conditions on $\varepsilon$ and $(\varepsilon_t)$ are, for example,
satisfied by $\varepsilon_{t} = \varepsilon t /(1000+t)$ and any
$\varepsilon\leq1/4$ with $\lambda= 1$
and $q = 2$. This spending
sequence $(\varepsilon_t)$ is the default spending sequence in the
R-package \textit{simctest}.

The conditions on $F$ are mild. For example, if
$F$ has a bounded density in a neighborhood of $\alpha$, then
$\xi=1$. If the distribution of $p$ is discrete and has finite support
(e.g., in a permutation test), then $\xi= 1$ if $\Prob[p=\alpha]=0$.
Otherwise, it is in principle possible to find $\alpha'>\alpha$ such that
\[
\beta= \Prob[p \leq\alpha] = \Prob\bigl[p \leq\alpha'\bigr],\qquad \Prob
\bigl[p=\alpha'\bigr] = 0.
\]
Applying the algorithm to $\alpha'$ instead of $\alpha$, we again have
$\xi=1$.

Henceforward the conditions of Theorem~\ref{thmfinstoptime} are
assumed to be satisfied with $\xi=1$. The algorithm will meet the
user-specified precision requirements with a finite expected effort if
it will terminate by time $\tau_{(N-2)}$ with probability one, or if
$\Prob[|\Int_{\tau_{(N-2)}}|> \Delta]=0$. As can be verified, with
$N-2$ of $N$ streams resolved the largest possible CI length occurs
when there are $\lfloor(N-2)/2 \rfloor$ positive outcomes. $N$ must
therefore satisfy $|I(\lfloor(N-2)/2\rfloor, \lceil(N-2)/2\rceil, 2;
\gamma)| \leq\Delta$. We shall call the minimal such $N$ the \emph
{blind minimal $N$, $N_{\Blind}$}.

\section{Choosing the number of streams} \label{secnostreams}
The computational effort of Algorithm~\ref{algbasic} can be large; see
Section~\ref{secsimulations}. In this section we introduce two
improvements concerning the choice of $N$: a \emph{pilot sample} that
can allow a smaller $N$ than $N_{\Blind}$, $N_{\Pilot}$, and an
estimate of the optimal $N \geq N_{\Pilot}$, using information from
the pilot.

\subsection{Reducing the simple minimum $N$}\label{secpilot}
Before running the main algorithm, we propose to first obtain a \emph
{pilot sample}, where $n$ streams are run and stopped at a maximum
number of steps $t_{\max}$, obtaining a preliminary confidence interval
$\Int_\Pilot=I(R_\Pilot, A_\Pilot, |\Unr_\Pilot|; \gamma_\Pilot)$,
where $I$ is defined in (\ref{defI}), $\gamma_\Pilot$ is some
pre-specified value (substantially) less than $\gamma$ and $R_\Pilot$,
$A_\Pilot$, $|\Unr_\Pilot|$ are the number of positive outcomes,
negative outcomes and unresolved streams.

In the main run the following interval can then be reported
%
\begin{equation}
\Int^{(\Pilot)}_t = I\bigl(R_t, A_t, |
\Unr_t|; \gamma- \gamma_\Pilot\bigr) \cap \Int_\Pilot.
\label{eqintersection}
\end{equation}
This respects the coverage probability $1-\gamma$, since a Bonferroni
correction was used.
We call the minimal $N$ such that for all $r\in\{0,1,\ldots,N-2\}:$
\[
\bigl|I(r, N-2-r, 2; \gamma-\gamma_{\Pilot})\cap\Int_\Pilot \bigr|\leq
\Delta
\]
the \emph{pilot-based minimal $N$} denoted by $N_{\Pilot}$. Given
$\Int_\Pilot$ it can be determined by a computational search.

\begin{figure}

\includegraphics{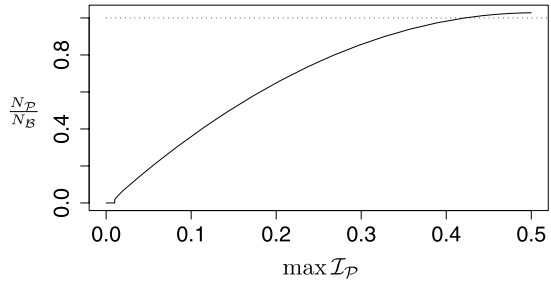}

\caption{Ratio of the pilot-based minimum $N$, $N_{\Pilot}$, over the
blind version, $N_{\Blind}$ as a function of the
rightmost point $\max I_{\Pilot}$ of the pilot sample interval, with
$\Delta= 0.01$, $\varepsilon= 0.0001$, $\gamma= 0.01$, $\gamma_\Pilot=
\gamma/10$. Here, $N_{\Blind}=68\mbox{,}311$.}\label{figminN}
\end{figure}
%
%

For $N\geq N_{\Pilot}$ the confidence interval will always reach
the desired length if at most 2 streams are unresolved.
$N_{\Pilot}$ can be much smaller than
$N_{\Blind}$. Indeed, after $N-2$ of $N$ streams in the main run are
resolved, the maximum CI length achievable is for a number of positive
outcomes $r$ that satisfies $r/(N-2) \in\Int_\Pilot$. As
demonstrated for pilot intervals $\Int_{\Pilot}$ to the left of 0.5 in
Figure~\ref{figminN}, the minimum number of streams needed
in the main run can be reduced substantially, in particular, if
$\Int_\Pilot$ lies far to the left (or right) of $0.5$.

Heuristically, the disadvantage of a small increase in the coverage
probability from $1-\gamma$ to $1-(\gamma-\gamma_{\Pilot})$ can be
outweighed by being able to exclude large intervals centered around 0.5.

\subsection{Approximation of the optimal number of streams}
\label{secoptstreams}
In this section, we choose $N$ within the range of possible
$N$s ($N \geq N_\Pilot$) in order to minimize $\E(e)$, where $e$
is defined in (\ref{eqeffort}). We use a heuristic
approach, which we only sketch briefly. Details can be found
in the supplementary material~\cite{supplement}.

From the pilot sample, we obtain an estimate of the probability of a
stream stopping before $t_{\max}$, its expected stopping time under
this event, and a preliminary estimate of $\beta$.

The expected stopping time of streams finishing after $t_{\max}$ is
predicted on the basis of the approximation $\Prob[\tau_i>t | \tau_i>t_{\max}] \approx c\sqrt{\log(t)/t}$.
This appears to be appropriate
(for a large enough $t_{\max}$) when the $p$-value distribution is
sufficiently ``well behaved'' around $\alpha$.

Using these quantities we can approximate the expected effort for
each $N$. The optimum $N$, denoted by $N_\Opt$, is found by searching
over a sensible range
$N_\Pilot\leq N \leq N_{\max}$.

\section{Stopping based on joint information}
\label{secseqhyp}
We now describe a testing procedure that analyzes the current set of
unresolved streams as a whole and allows the algorithm to stop with
more unresolved streams. It reports a lower bound $r_t$ ($a_t$) on the
number of positive (negative) outcomes from\vadjust{\goodbreak} the remaining streams if
both of the following hypotheses are rejected,
\[
H^+_0\dvtx \bigl|\{i\in\Unr_t\dvtx p_i \leq
\alpha\}\bigr|< r_t,\qquad  H^-_0\dvtx \bigl|\{i\in\Unr_t\dvtx
p_i>\alpha\}\bigr| < a_t,
\]
where $r_t, a_t \geq0$ and $r_t +a_t \leq|\Unr_t|$. The choice of
$r_t$ and $a_t$ is discussed later.

The hypotheses will be rejected for large values of the test statistics,
\[
T^+ = \sum^{|\Unr_t|}_{i=r_t} \Ind
\bigl[G_t^\alpha\bigl(S_t^{(i)}\bigr)
\leq\eta\bigr], \qquad T^- = \sum^{|\Unr_t|-a_t+1}_{i=1} \Ind
\bigl[G_t^\alpha\bigl(S_t^{(i)}\bigr)
\geq 1-\eta\bigr],
\]
where $S_t^{(1)}\leq\cdots\leq S_t^{(|\Unr_t|)}$ are the \emph{ordered}
partial sums corresponding to the unresolved streams, $\eta$ is a
chosen (small) positive value and
\[
G_t^\alpha(x) = \Prob_\alpha[S_t \leq x
| \tau> t],
\]
that is, $G_t^\alpha$ is the CDF of a cumulative sum of $t$ Bernoulli
variables with success probability $\alpha$, conditional on not having
hit either boundary by time~$t$. This function is computed recursively.

The random variable $X$ is said to be \emph{stochastically smaller}
than the random variable
$Y$, denoted $X
\leq_{st} Y$, if $\Prob(X\leq x)\geq\Prob(Y\leq x)$ for all $x \in
\R$.

\begin{thmm}\label{thmtest}
Under $H^+_0$, $T^+ \leq_{st} B^+$ and under $H_0^-$, $T^- \leq_{st}
B^-$, where $B^+$ and $B^-$ are Binomial variables with success
probability $\eta$ and size $|\Unr_t| - r_t + 1$ and $|\Unr_t| - a_t +
1$, respectively.
\end{thmm}
$H_0^+$ and $H_0^-$ can therefore be rejected \emph{conservatively}
when $T^+$ and $T^-$ are significantly large for the corresponding
Binomial variables.

Using Bonferroni correction, a minimum coverage probability of
$1-\gamma
$ is guaranteed if for all $t$ we compute a confidence interval
\[
\Int^{\Joint}_t = I\bigl(\tilde{R}_t,
\tilde{A}_t, |\tilde{\Unr}_t|; \gamma -
\gamma_\Pilot- \gamma_\Joint\bigr) \cap\Int_\Pilot,
\]
where $(\tilde{R}_t, \tilde{A}_t, |\tilde{\Unr}_t|) = (R_t+r_t,
A_t+a_t, |\Unr_t| - r_t - a_t)$ if the test rejects, $(R_t, A_t,\break
|\Unr_t|)$ otherwise, and $\gamma_\Joint< \gamma- \gamma_\Pilot$ is
an upper bound on the overall probability of wrongly rejecting either
hypothesis at any point in time. To guarantee this bound,
at each time $t$, each hypothesis is tested at level $\xi_t/2$,
where
$\xi_1, \xi_2,\ldots\geq0$ are constants satisfying
$\sum_{i=1}^{\infty} \xi_{i} = \gamma_\Joint$.

$r_t$ and $a_t$ are chosen such that $|\Int^{\Joint}_t| \leq\Delta$ if
both tests reject, so that the algorithm can stop immediately if this occurs.

The procedure is mostly useful when the number of resolutions required,
$r_t + a_t$, is small compared to the number of remaining streams
$|\Unr_t|$. As an extreme example, suppose that $r_t = 1$, $a_t = 0$ and
$|\Unr_t| = 100$. In this case, it can be possible to conclude with
virtual certainty that \emph{at least} 1 of the 100 streams has a
$p$-value less than $\alpha$, when concluding the same about any
individual stream could require many more samples.

In this procedure there are a number of free parameters that we set
somewhat heuristically. From a small simulation study we established
that choosing $\eta=0.05$ gave good results. As for $r_t$ and $a_t$,
they are chosen to be equal and then as small as possible subject to
the algorithm terminating if the hypotheses can be rejected, since for
simple $p$-value distributions it is likely that the unresolved
$p$-values would be roughly evenly distributed around $\alpha$.

In the simulation studies that follow and in the R-implementation,
$\gamma_\Joint= \gamma/10$, $\xi_t$ is only positive when $t = t_i =
2i \times10^5$ for $i \in\N$ and $\sum_1^{t_i} \xi_t= \gamma_\Joint
\times20/(20+i)$.

\section{Simulations}
\label{secsimulations}
This simulation study illustrates the effort required by our algorithm
and the effect of the improvements in Sections~\ref{secnostreams} and
\ref{secseqhyp}. For all experiments we set $\alpha= 0.05$, $\Delta
= 0.02$, $1-\gamma= 0.99$, $\varepsilon= 0.0001$ and $\varepsilon_t =
\varepsilon1000/(1000+t)$. Four $p$-value distributions were considered,
$\mathrm{Beta}(1,x)$ with $x$ such that $\Prob[p\leq\alpha] =
\alpha,
0.7, 0.9, 0.99$, that is, $x=1$ (a uniform distribution) and roughly
$x=23.5$, $x=44.9$ and $x=89.8$, respectively. As before, the
effort is measured by the total number of samples generated.

Table~\ref{tabeffort} shows the average effort based on 100
replicated runs in the left subcolumns. In the right subcolumns we
report the estimated standard error of the corresponding estimate,
that is, the standard deviation of the sample divided by $\sqrt{100}$.

\begin{table}[b]
\caption{Average effort (in millions) of our adaptive methods (``No
test'' and ``With test'') compared with the minimum $N$ and the optimal $N$}
\label{tabeffort}
\begin{tabular*}{\textwidth}{@{\extracolsep{\fill}}lcccccccc@{}}
\hline
&\multicolumn{2}{c}{$\bolds{\beta=0.05}$}&\multicolumn
{2}{c}{$\bolds{\beta= 0.7}$}&\multicolumn{2}{c}{$\bolds{\beta= 0.9}$}&\multicolumn
{2}{c@{}}{$\bolds{\beta= 0.99}$}\\[-6pt]
&\multicolumn{2}{c}{\hrulefill}&\multicolumn
{2}{c}{\hrulefill}&\multicolumn{2}{c}{\hrulefill}&\multicolumn{2}{c@{}}{\hrulefill}\\
&\textbf{Av.} & \textbf{(S.E.)}&\textbf{Av.} & \textbf{(S.E.)}&\textbf{Av.} & \textbf{(S.E.)}&\textbf{Av.} & \textbf{(S.E.)}\\
\hline
Optimal $N$ &12.3&(0.14)&3329&\phantom{0}(35)&539&(8.4)&16.2&(0.08)\\
Min. $N$ &12.5&(0.16)&8498&(296)&548&(9.2)&16.1&(0.08)\\
No test &10.5&(0.22)&3324&\phantom{0}(41)&568&(7.9)&10.4&(0.10)\\
With test &\phantom{0}8.0&(0.19)&1541&\phantom{0}(18)&317&(5.2)&10.4&(0.09)\\
\hline
\end{tabular*}
\end{table}

The first two rows report the average effort for the optimal $N$ (which
would not be available in practice) and the minimum $N$, $N_\Blind$,
when using Algorithm~\ref{algbasic} without any of the improvements
suggested in Sections~\ref{secnostreams} and~\ref{secseqhyp}. These
were computed by resampling from $10^6$ pre-simulated replicates of the
tuple (stopping-time, outcome), for each distribution, from which we
emulated the operation of the algorithm. (Finding the optimal $N$ would
otherwise have taken too much time.)

The third row illustrates the improvements of Section \ref
{secnostreams}, which concern the choice of $N$, setting $\gamma_\Pilot= 0.1 \gamma$.
In the fourth row we additionally implemented
the test on joint information, described in Section~\ref{secseqhyp},
with $\gamma_\Joint= 0.1 \gamma$. In both of these rows each value
represents the average effort observed from actually running the
algorithm 100 times. Each run used its own pilot sample consisting of
1000 streams forced to terminate after 1000 steps. The effort of the
pilot is included in the average effort.

First consider the difference between the third and fourth rows of
Table~\ref{tabeffort}. The testing procedure can reduce the effort
substantially, namely by 24\%, 54\%, 44\% in the first three cases,
although in the last case the reduction is not significant.

For the Uniform and Beta distribution with power 0.99, the optimal $N$
and $N_\Blind$ turn out to be equal. Hence, the reduction of the effort
seen in the third row over the first two rows is mostly due to the
intersection method described in Section~\ref{secpilot}, which has
allowed a smaller choice of $N$, $N_\Pilot$.

For the Beta distribution with power 70\%, the effort for the minimal
$N$, in the second row, is over 2.5 times larger than for the optimal
$N$, in the first row. As result, in this example it was crucial to
estimate this optimum, by the procedure described in Section \ref
{secoptstreams}. The difference between the effort for the optimal $N$
(unknown in practice) and the adaptively chosen $N_\Opt$ is not
significant (although in this example enough simulations would show
that the optimal $N$ still performed better). As previously mentioned,
introducing the testing procedure in this example further reduces the
effort by a considerable margin, as demonstrated in the fourth row. It
is of some comfort that the best improvements from the methodology of
Sections~\ref{secnostreams} and~\ref{secseqhyp} were found in the
computationally most demanding scenario.

In the third row, for the Beta distribution with power 90\%, adaptively
choosing $N$ actually increased the effort, although not substantially.
The average $N_\Opt$ chosen is roughly 10,000, whereas $N_\Blind$ in the
second row is 17,055 (for this distribution it is also the optimal $N$).
We would expect to reduce the effort on this basis. However, this does
not appear to completely compensate for the effort of the pilot and the
error in coverage probability lost in computing the pilot-based CI.
However, with the test we reduce the effort by 40\% and improve on both
efforts reported in the first two rows for this distribution.

Overall, from these experiments it seems that our suggested
improvements reduce the expected effort substantially, as is best
summarized in the difference between the bottom row and either of the
first two.

For future reference, the default settings of our algorithm are those
of the bottom row, namely: $\varepsilon= \Delta/200$, $\varepsilon_t =
\varepsilon1000/(1000+t)$, $\gamma_\Pilot= \gamma_\Joint= 0.1 \gamma$
and a pilot sample of 1000 streams terminated at $t_{\max}=1000$.

\section{Adaptive CI length}\label{secadaptci}
When one resampling step is computationally demanding, the expected
efforts listed in Table~\ref{tabeffort} may appear prohibitive. In
this case, we recommend relaxing the fixed requirements on $\Delta$,
that is, allowing $\Delta$ to depend on the ``location'' of the
confidence interval. This can reduce the expected effort of the
algorithm substantially.\vadjust{\goodbreak}

As a rule of thumb, the closer the power is to 0.5 the higher the
expected effort (compare, e.g., the efforts for $\beta=0.05$ and
$\beta=0.7$ in Table~\ref{tabeffort}): first, because the $p$-value
distribution tends to have more mass around $\alpha$, meaning that each
stream in the algorithm has a higher expected running-time, and second
because the length of the confidence interval is largest when there are
the same number of positive and negative outcomes.

On the other hand, we anticipate that if the power is around 0.5 or for
that matter anywhere in the interval $[0.1, 0.9]$, say, the user will
often only require a small enough confidence interval to conclude that
$\beta$ is not close $\alpha$ or 1. Indeed, a~typical reason why one
needs the power of a test is to check that the probability of rejection
under the null hypothesis is close to $\alpha$ (which is typically
small) or that under an alternative hypothesis $\beta$ is close to 1.

Let $C = \{\beta\in[0,1]^2\dvtx \beta_1 \leq\beta_2\}$ denote the set of
all possible confidence intervals for $\beta$. We allow the analyst to
pre-specify a subset of $C$, $A$, say, such that if the current
confidence interval is an element of $A$ the algorithm terminates immediately.

It is reasonable to enforce that $A$ satisfy the following three properties:
\begin{longlist}[(iii)]
\item[(i)] $A$ is closed.
\item[(ii)] $\{(\beta, \beta)^T\dvtx \beta\in[0,1]\} \subseteq A$ (CIs of
length 0 are allowed).
\item[(iii)] $\forall\beta\in A\dvtx \forall\alpha\in C\dvtx \beta_1 \leq
\alpha_1 \leq\alpha_2 \leq\beta_2 \Rightarrow\alpha\in A$ (a
subinterval of an allowed CI is allowed).
\end{longlist}
The next result shows that specifying $A$ is equivalent to specifying
the maximum CI length allowed as a function of the CI's midpoint.
%
\begin{lem}\label{lemmidpoint}
If $A \subseteq C$ satisfies \textup{(i)--(iii)}, then there exists a function
$\Delta\dvtx [0,1]\to[0,1]$ such that
for all $\beta\in C\dvtx \beta\in A \Leftrightarrow\beta_2 - \beta_1
\leq\Delta(\frac{\beta_1 + \beta_2}{2})$.
\end{lem}
All of the theory we have presented in Sections~\ref{secmain}--\ref
{secseqhyp} can be incorporated unaltered into an algorithm with
adaptive $\Delta$, with the single exception that finding $N_\Pilot$
requires a brute-force search---one must ensure that $\Delta(M)$ will
be met after $N-2$ streams have stopped, for any possible CI midpoint
$M$ arising from all the possible outcomes of $N-2$ streams.

\begin{figure}[b]
\vspace*{-2pt}
\includegraphics{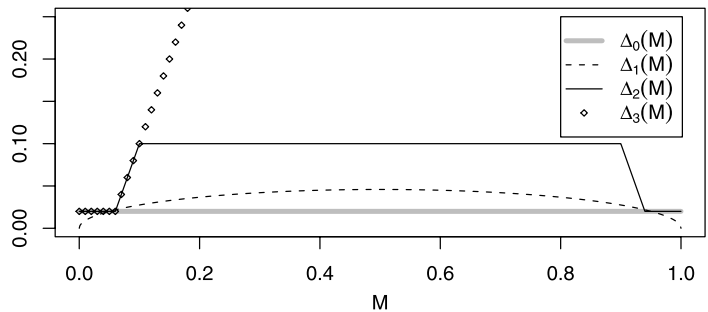}

\caption{The four midpoint functions $\Delta_i$ used in Table
\protect\ref
{tabaeffort}.}\label{figdeltamid}
\end{figure}

The effort of our recommended method for fixed $\Delta$ is repeated
from the fourth row of Table~\ref{tabeffort} to the first row of Table
\ref{tabaeffort}. These results are equivalent to a case where for all
$M \in[0,1]$, $\Delta(M) = 0.02 = \Delta_0(M)$. In the next rows of
Table~\ref{tabaeffort} we present the average effort of the algorithm
for three other functions of the midpoint, all of which are illustrated
in Figure~\ref{figdeltamid}. Depending on what is easiest to present,
the rule is described through $\Delta$ or by the equivalent $A$.
\begin{longlist}[(1)]
\item[(1)] $\Delta_1(M) = 0.02\sqrt{M (1-M)}/(\sqrt{0.05 \cdot0.95})$.
A function that allows\break roughly the same number of streams to remain
unresolved for any $\beta$.\vadjust{\goodbreak} Because the CI midpoint cannot be 0 or 1
exactly the fact that $\Delta(0)=\Delta(1)=0$ is not problematic.
\item[(2)]
$A_2$ is the largest set of confidence intervals that satisfies
(i)--(iii) and
that satisfies $\forall\beta\in A_2\dvtx \beta_2 - \beta_1 \leq0.1$ and
$ \forall\beta\in A_2$ with ($\beta_1 \leq0.05$ or $\beta_2 \geq
0.95$): $\beta_2 - \beta_1 \leq0.02$---a CI length of 0.02 is needed
for high or low powers, but a CI length of 0.1 is admissible otherwise.
\item[(3)]
$A_3$ is the largest set of confidence intervals that satisfies
(i)--(iii) and
that satisfies $ \forall\beta\in A_3$ with $\beta_1 \leq0.05$:
$\beta_2 - \beta_1 \leq0.02$. A precise estimate is only required if the
confidence interval is at least partly to the left of $\alpha$ and any
interval is admissible otherwise.\vspace*{-2pt}
\end{longlist}

\begin{table}
\caption{Average effort (in millions) for different functions of the CI
midpoint}\label{tabaeffort}
\begin{tabular*}{\textwidth}{@{\extracolsep{\fill}}lcccccccc@{}}
\hline
&\multicolumn{2}{c}{$\bolds{\beta=0.05}$}&\multicolumn
{2}{c}{$\bolds{\beta= 0.7}$}&\multicolumn{2}{c}{$\bolds{\beta= 0.9}$}&\multicolumn
{2}{c@{}}{$\bolds{\beta= 0.99}$}\\[-6pt]
&\multicolumn{2}{c}{\hrulefill}&\multicolumn
{2}{c}{\hrulefill}&\multicolumn{2}{c}{\hrulefill}&\multicolumn{2}{c@{}}{\hrulefill}\\
\textbf{Function}&\textbf{Av.} & \textbf{(S.E.)}&\textbf{Av.} & \textbf{(S.E.)}&\textbf{Av.} & \textbf{(S.E.)}&\textbf{Av.} & \textbf{(S.E.)}\\
\hline
$\Delta_0$&8.0&(0.19)&1541&(18)&317&(5.2)&10.4&(0.09)\\
$\Delta_1$&7.8&(0.20)&185&(3.2)&131&(2.3)&26.2&(0.77)\\
$\Delta_2$&8.4&(0.46)&17.1&(0.46)&9.0&(0.06)&5.5&(0.08)\\
$\Delta_3$&8.4&(0.46)&0.7&($<$0.01)&0.6&($<$0.01)&0.5&($<$0.01)\\
\hline
\end{tabular*}   \vspace*{-2pt}
\end{table}

For the Uniform distribution, since all rules have $\Delta(0.05) =
0.02$, we would expect the effort to be comparable, as is observed. On
the other hand, we see a dramatic reduction of the effort in other
columns where the rule has allowed less precision. Overall, if we
consider for example the effort for $\Delta_2$, we hope that with this
compromise the algorithm can be used in practice for moderately
complicated tests.\vspace*{-2pt}

\section{Example: Permutation test}\label{secpermtest}
Using exactly the example of \cite{boos00}, we computed the power of a
permutation test on the difference of the means of two Gaussian
samples, with sizes $K=4$ and $L=8$, identical standard deviation
$\sigma$ and\vadjust{\goodbreak} standardized differences $(\mu_\mG-\mu_\mC)/\sigma
=0.5, 1,
1.5$ and $2$. We used a fixed $\Delta=0.01$ and coverage probability
$0.99$. Our other parameters were set to the defaults listed at the end
of Section~\ref{secsimulations}.

The results are presented in Table~\ref{tabpermtest}. In three of the
four cases our confidence interval excludes the corresponding estimate
in \cite{boos00} (although not after adding or subtracting two of their
standard errors). Of course, our computational effort is considerably
larger---but our key contribution is in providing a mechanism that
\emph{guarantees} the precision of the result.

In this simple example it is in fact possible to compute the $p$-value
of each dataset exactly by evaluating all 495 permutations. Because of
this the power can be estimated by standard methodology with a
Binomial-based confidence interval. In each case, a very accurate
estimate of $\beta$ was obtained by generating $10^6$ datasets and
computing the $p$-value for each exactly. The resulting estimates are
presented in the first row of Table~\ref{tabpermtest}, using the
convention ${_ax_b}$ to mean that the estimate is $x$, and the
confidence interval is $[a,b]$. In the second row we present the
results of our algorithm, using a fixed $\Delta=0.01$ and coverage
probability 0.99. In all cases, the ``true'' power falls within our
estimated confidence interval, as would be expected.
For the convenience of the reader, the third row presents the estimated
powers and standard errors computed in \cite{boos00}.\vspace*{-2pt}

\begin{table}
\caption{Power of the permutation test for the difference of means}
\label{tabpermtest}\vspace*{-2pt}
\begin{tabular*}{\textwidth}{@{\extracolsep{\fill}}lcccc@{}}
\hline
$\bolds{\Delta/\sigma}$ & \textbf{0.5} & \textbf{1.0} & \textbf{1.5} & \textbf{2.0}\\
\hline
Truth
&${_{0.183}0.184_{0.185}}$&${_{0.441}0.442_{0.443}}$&${_{0.728}0.729_{0.730}}$&${_{0.912}0.912_{0.913}}$\\
Our method
&${_{0.182}0.185_{0.192}}$&${_{0.440}0.443_{0.450}}$&${_{0.726}0.729_{0.736}}$&${_{0.910}0.914_{0.920}}$\\
Boos and Zhang &$0.175\ (0.006)$ & $0.439\ (0.008)$ & $0.731\ (0.007)$ &
$0.921\ (0.005)$\\
\hline
\end{tabular*}    \vspace*{-2pt}
\end{table}

\section{Conclusions}
\label{secconclusions}
We have proposed an open-ended algorithm to compute a conservative
confidence interval for $\beta$, (almost) without any assumptions on
the distribution of the $p$-value (Theorem~\ref{thmfinstoptime}). In
practice, the method can be computationally expensive. However,
various improvements (Sections~\ref{secnostreams} and
\ref{secseqhyp}) reduce the computational effort for fixed $\Delta$ by
a sizeable margin. An adaptive $\Delta$ (Section~\ref{secadaptci}) can
ensure that the effort is only large if the estimated power is in a
region where a high precision is required.

There remain areas of potential improvement: for instance the balance
between the error spent on $\varepsilon$, the pilot and the testing
procedure could be explored in more depth, as well as the choice of the
spending sequences $\varepsilon_{t}$ and $\xi_{t}$. The test for stopping
based on joint information in Section~\ref{secseqhyp} is somewhat
ad-hoc, and conceivably a more powerful test could be derived. Finally,
of course, the computational effort could also potentially be reduced
by making additional assumptions on the $p$-value distribution.

How conservative is the confidence interval? From a few simple
experiments, we found the length to\vadjust{\goodbreak} be roughly twice as large as it
needs to be for the nominal coverage probability. Although we have been
conservative in many aspects of the algorithm, this disparity appears
to be almost entirely due to the contribution from unresolved streams
in (\ref{defI}). This is effectively the price of making almost no
assumptions on the distribution of the $p$-values.

\begin{appendix}\label{app}
\section{Finite expected stopping time}
The proof of Theorem~\ref{thmfinstoptime} requires preliminary lemmas
and the following recursive definition of the stopping boundaries
from \cite{gandy09}:
%
\begin{eqnarray}
\label{equldef} %
U_t &=& \min\bigl\{j \in\N\dvtx
\Prob_{\alpha} ( \tau\geq t, S_t \geq j) + \Prob_{\alpha}(
\tau< t, S_\tau\geq U_\tau) \leq\varepsilon_t
\bigr\},
\nonumber
\\[-8pt]
\\[-8pt]
\nonumber
L_t &=&\max\bigl\{j \in\Z\dvtx \Prob_{\alpha}( \tau\geq t,
S_t \leq j) + \Prob_{\alpha}(\tau< t,S_\tau\leq
L_{\tau} ) \leq\varepsilon_t\bigr\}. %
\end{eqnarray}

\begin{lem}\label{lem1}
If there exist constants $\lambda> 0$, $q>0$ and $T \in\N$ such
that $\varepsilon_t - \varepsilon_{t-1} \geq\lambda t^{-q}$ for all $t
\geq T$, then, for all $t \geq T$,
\begin{eqnarray*}
U_t \leq \bigl\lceil t \alpha+ \sqrt{t(q \log t - \log \lambda)/2}
\bigr\rceil,\qquad
L_t \geq \bigl\lfloor t \alpha- \sqrt{t (q \log t - \log\lambda )/2}
\bigr\rfloor.
\end{eqnarray*}
\end{lem}

The square root is well defined since $1 \geq\varepsilon_t -
\varepsilon_{t-1} \geq\lambda t^{-q}$.

\begin{pf}
We will show $ \Prob_{\alpha}(\tau\geq t, S_t \geq U_t^*) +
\Prob_{\alpha}(\tau<t, S_{\tau} \geq U_{\tau}) \leq\varepsilon_t$ for
\mbox{$t\geq T$}.
By (\ref{equldef}) this implies $U_t\leq \lceil t \alpha+
\sqrt{t (q \log t - \log\lambda)/2}  \rceil=:U_t^\ast$.

First,
(\ref{equldef}) implies
\begin{eqnarray*}
\Prob_{  \alpha}(\tau< t, S_{\tau} \geq U_{\tau})& =&
\Prob_{  \alpha}(\tau \geq t-1, S_{t-1} \geq U_{t-1})+
\Prob_{  \alpha}(\tau <t-1, S_{\tau} \geq U_{\tau
})\\
&\leq&
\varepsilon_{t-1}.
\end{eqnarray*}
Furthermore, by Hoeffding's inequality \cite{hoeffding63},
\begin{eqnarray*}
\Prob_{ \alpha}\bigl(\tau\geq t, S_t \geq U_t^*
\bigr)&\leq &\Prob_{ \alpha}\bigl(S_t \geq U_t^*\bigr)
= \Prob_{ \alpha}\bigl(S_t/t - \alpha \geq U_t^*/t -
\alpha\bigr)
\\
& \leq&\exp\bigl\{-2t\bigl(U_t^*/t-\alpha\bigr)^2\bigr\}
\leq\lambda t^{-q} \leq\varepsilon_t - \varepsilon_{t-1},
\end{eqnarray*}
finishing the proof of $U_t\leq U_t^\ast$.
The bound for $L_t$ can be shown similarly.
\end{pf}

The above formally confirms the observation in \cite{gandy09}, main text, page
1507 and Figure 4, that $U_t - L_t$ appears to be proportional
to $\sqrt{t \log t}$ for large $t$. Indeed, the spending sequence used,
$\varepsilon_t = \varepsilon t/(1000+t)$, satisfies the conditions of the
lemma with $\lambda=1$ and $q=2$ (if one chooses $\varepsilon\leq1/4$).
%
\begin{lem}\label{lem2}
Suppose that $F$ is H\"older continuous with exponent $\xi$ in a
neighborhood of $\alpha$, that the conditions of Lemma~\ref{lem1} hold,
and that $\varepsilon\leq1/4$. Then, for any $\eta\in(0,1)$, there
exist constants $\kappa$ and $\tilde{T}$ such that
\[
\Prob(\tau> t) \leq2 \mathrm{e}^{-2t^{\eta}} + \kappa t^{\xi
(\eta- 1)/2}, \qquad t \geq
\tilde{T}.
\]
Hence,
$\Prob(\tau> t) = o(t^{d})$ for any $d > -\xi/2$.\vadjust{\goodbreak}
\end{lem}

\begin{pf}
Let $F$ be the CDF of $p$. Then, for any $t \in\N$,
\[
\Prob(\tau> t) = I\bigl\{\bigl[0, p_t^-\bigr]\bigr\} + I\bigl\{
\bigl(p_t^-, p_t^+\bigr)\bigr\} + I\bigl\{
\bigl[p_t^+, 1\bigr]\bigr\},
\]
where $I\{A\} = \int_{A}\Prob_p(\tau>t) \,\rd F(p)$ and $0 \leq
p_t^-<\alpha< p_t^+ \leq1$. When $0 \leq p \leq p_t^-$ and $L_t/t -
p_t^- > 0$,
\[
\Prob_p(\tau> t) \leq\Prob_p(S_t >
L_t) \leq\Prob_{p_t^-}(S_t > L_t)
\leq\exp\bigl\{-2t\bigl(L_t/t - p_t^-\bigr)^2
\bigr\},
\]
using Hoeffding's inequality for the rightmost bound.
Hence, letting
\[
p_t^- = \max\bigl\{L_t/t - t^{(\eta- 1)/2}, 0\bigr\},\qquad  t
\in\N
\]
for some $\eta\in\R$, we get
\[
\Prob_p(\tau> t) \leq\exp\bigl\{-2 t^\eta\bigr\},\qquad  0 \leq
p \leq p_t^-, t\in\N.
\]
Do we have $0 \leq p_t^- < \alpha$? The lower bound is obvious. The
upper bound also holds, since the proof of Theorem 2 in \cite{gandy09}
shows that if $\varepsilon\leq1/4$, then $L_t/t < \alpha$ for all $t
\in
\N$.

Similarly we can define
$p_t^+ = \min\{U_t/t + t^{(\eta-1)/2}, 1\}$, $t \in\N$,
guaranteeing that $\alpha< p_t^+ \leq1$. Then, for any $\eta\in\R$,
\[
\Prob_p(\tau> t) \leq\exp\bigl(-2t^{\eta}\bigr),\qquad
p_t^+ \leq p \leq1, t \in\N.
\]
We therefore have
%
\begin{equation}
I\bigl\{\bigl[0, p_t^-\bigr]\bigr\} + I\bigl\{\bigl[p_t^+,
1\bigr]\bigr\} \leq2 \exp\bigl(-2 t^{\eta}\bigr). \label
{eqwithoutnbhood}
\end{equation}
It remains for us to obtain a bound on $I\{(p_t^-, p_t^+)\}$. Using
Theorem 1 in \cite{gandy09}, $U_t - \alpha t = o(t)$, $\alpha t - L_t =
o(t)$. Thus, by restricting $\eta<1$, $p_t^- \rightarrow\alpha$,
$p_t^+ \rightarrow\alpha$ and there exists a time $T^*$ such that $F$
is H\"older continuous over $(p_t^-, p_t^+)$ for all $t \geq T^*$. It
follows that for some constant $h>0$,
\[
I\bigl\{\bigl(p_t^-, p_t^+\bigr)\bigr\} \leq\int
_{(p_t^-, p_t^+)} \,\rd F(p) \leq F\bigl(p_t^+\bigr) - F
\bigl(p_t^-\bigr) \leq h\bigl(p_t^+ - p_t^-
\bigr)^{\xi},\qquad t \geq T^*.
\]
Let $\tilde{T} = \max\{T, T^*, 2\}$, where $T$ is defined in Lemma
\ref
{lem1}. For $t \geq\tilde{T}$,
\begin{eqnarray*}
I\bigl\{\bigl(p_t^-, p_t^+\bigr)\bigr\} &\leq& h
\bigl(p_t^+ - p_t^-\bigr)^{\xi}
\\
&\leq& h \bigl[2 t^{(\eta- 1)/2} + 2 \bigl[\sqrt{t(q \log t - \log\lambda )/2} +
1\bigr]/t \bigr]^{\xi}
\\
&\leq& h \bigl[2 t^{(\eta- 1)/2} + 2 \bigl[\sqrt{(q+a)/2} \sqrt{t\log t} + 1
\bigr]/t \bigr]^{\xi}
\\
&\leq &h \bigl[2 t^{(\eta- 1)/2} + b \sqrt{\log t/t} \bigr]^{\xi}
\\
& \leq & h \bigl[(2+c)t^{(\eta- 1)/2} \bigr]^{\xi}\qquad \mbox{(requiring $
\eta> 0$),}
\end{eqnarray*}
where $a = \max\{0, -\log\lambda/\log\tilde{T}\}$,
$b = 2 (\sqrt{(q+a)/2}+1)$,
$c = b \sqrt{\log t/t}|_{t = \tilde{T}}$.
We needed $\tilde{T}\geq2$ in the definition of $a$ and used it in the
third inequality ($1 < \sqrt{2 \log2}$). Using (\ref
{eqwithoutnbhood}), the proof is complete after we take $\kappa= h (2
+ c)^{\xi}$.~%
\end{pf}

\begin{pf*}{Proof of Theorem~\ref{thmfinstoptime}}
Using standard results for order statistics~\cite{embrechts97},
\begin{eqnarray*}
\Prob(\tau_{(N-k)} > t) = \sum_{j=0}^{N - k - 1}
\pmatrix{{N}
\cr
{j}} \Prob(\tau> t)^{N-j} \Prob(\tau\leq
t)^j \leq c_1 \Prob(\tau>t)^{k+1}
\end{eqnarray*}
for $t\geq0$ and some $c_1 > 0$. Therefore, using Lemma~\ref{lem2}
\begin{eqnarray*}
\E(\tau_{(N-q)}) = \sum_{t = 0}^{\infty}
\Prob(\tau_{(N-k)} > t) \leq1+ \sum_{t = 1}^{\infty}
c_1 \Prob(\tau> t)^{k+1} \leq1+ \sum
_{t = 1}^{\infty} c_2 t^{(k+1)d}
\end{eqnarray*}
for all $d > -\xi/2$, with $c_2$ chosen based on $c_1$ and $d$. The
summation in the right-hand side is finite if the exponent of $t$ is
strictly smaller than $-1$. $\lfloor2/\xi\rfloor$ is the smallest
possibility for $k \in\N$ such that there exists a $d>-\xi/2$ with
$(k+1)d <-1$.
\end{pf*}

\section{Hypothesis test}
The proof of Theorem~\ref{thmtest} first requires the following lemma.
%
\begin{lem}\label{lemstochbound}
Suppose that $X_j^1$ and $X_j^2$ are two sequences of independent
Bernoulli variables with success probabilities $\pi_1$ and $\pi_2$,
respectively, where $0\leq\pi_1 \leq\pi_2 \leq1$, and put $S_t^{k} =
\sum_{j=1}^t X_t^k$ for $k = 1,2$. Let $\{l_t\dvtx t \in\N\}$ and $\{u_t\dvtx t \in\N\}$ be two arbitrary integer sequences, and let
\[
\tau_k = %
\cases{ \infty,& \quad $\mbox{if $l_t
<S^k_t < u_t$ for all $t \in\N $}$,
\vspace*{2pt}
\cr
\min\bigl\{j\dvtx S_j^k \leq
l_j \mbox{ or } S_j^k \geq
u_j\bigr\}, & \quad $\mbox{otherwise}.$ } %
\]
Then if $\Prob[\tau_k > t]>0$ for $k = 1,2$,
\[
\bigl[S^1_{t}|\tau_1 > t\bigr]
\leq_{st}\bigl[S^2_{t}|\tau_2 > t
\bigr].
\]
\end{lem}

\begin{pf}
We will require a stronger form of stochastic ordering: for two
discrete RVs $X$ and $Y$, $X$ is smaller than $Y$ with respect to the
likelihood ratio order, denoted $X \leq_{lr} Y$, if
%
\begin{equation}
\frac{f_{X}(x)}{f_{Y}(x)} \downarrow x\qquad \mbox{on the support set of $Y$,}
\label{eqlrtorder}
\end{equation}
where $f_X$ and $f_Y$ are the probability mass functions (PMFs) of $X$
and $Y$ \cite{keilson82}, page~184. Further, a discrete RV $Z$ has a
log-concave distribution if \cite{keilson71}
%
\begin{equation}
f_Z(x)^2 \geq f_Z(x-1)f_Z(x+1),\qquad
x \in\N. \label{eqlconcav}
\end{equation}
$[S^1_1|\tau_1 > 1]$ and $[S^2_1|\tau_2 > 1]$ have log-concave
distributions and $[S^1_1|\tau_1 > 1] = X^1_{1} \leq_{lr} X^2_1 =
[S^2_1|\tau_1 > 1]$. Suppose the same holds true for $[S^1_t|\tau_1 >
t]$ and $[S^2_{t}|\tau_2 > t]$. For $k=1,2$, $[S^k_{t+1}|\tau_k > t] =
[S^k_{t}|\tau_k > t] + X^k_{t+1}$ is a convolution of two random
variables with log-concave distributions, implying that it has itself a
log-concave distribution\vadjust{\goodbreak} \cite{keilson71}, Lemma page~387. Using \cite{keilson71}, Theorem~2.1(d)
\begin{eqnarray*}
\bigl[S^1_{t+1}|\tau_1 > t\bigr] &=&
\bigl[S^1_{t}|\tau_1 > t\bigr] +
X^1_{t+1} \leq_{lr} \bigl[S^2_{t}|
\tau_2 > t\bigr] + X^1_{t+1}
\\
&\leq_{lr} &\bigl[S^2_{t}|\tau_2 >
t\bigr] + X^2_{t+1} = \bigl[S^2_{t+1}|
\tau_2 > t\bigr],
\end{eqnarray*}
using the properties assumed to be true at $t$ and the log-concavity,
likelihood ratio ordering and independence of $X^1_{t+1}$ and $X^2_{t+1}$.

For $k=1,2$, conditioning on $\tau_k > t+1$ restricts the support of
$[S^1_{t+1}|\tau_1 > t]$ and $[S^2_{t+1}|\tau_2 > t]$ to a same,
smaller set, and (where supported) the new PMF is the old multiplied by
a constant $c_k$. Therefore, directly from (\ref{eqlrtorder}) and
(\ref
{eqlconcav}), we conclude that $[S^1_{t+1}|\tau_1 > t+1] \leq_{lr}
[S^2_{t+1}|\tau_2 > t+1]$, and both distributions are log-concave.

By induction, these properties are true for all $t$. Likelihood ratio
order implies the usual stochastic order \cite{keilson82}, completing
the proof.
\end{pf}

\begin{pf*}{Proof of Theorem~\ref{thmtest}}
Let $n_t=|\Unr_t|$. $T^+$ can be bounded above by
\[
T^+ \leq\sum^{n_t}_{i=r_t} \Ind
\bigl[G_t^\alpha\bigl(\tilde{S}_t^{(i)}
\bigr) \leq \eta \bigr] = \tilde{T}^+,
\]
where $\{\tilde{S}_t^{(i)}\dvtx i=r_t, \ldots, n_t\}$ are the partial sums
corresponding to $p_{(r_t)} \leq p_{(r_t+1)} \leq\cdots\leq p_{(n_t)}$,
the largest ordered $p$-values of the unresolved streams.

Under $H_0^+$, $p_{(i)} > \alpha$ for $i = r_t, \ldots, n_t$. Let
$S^{\alpha}_t$ be a partial sum generated by a $p$-value equal to
$\alpha$ and let $\tau_{\alpha}$ denote its stopping-time. By Lemma
\ref
{lemstochbound},
\[
\bigl[S^{\alpha}_t | \tau_{\alpha} > t\bigr]
\leq_{st} \bigl[\tilde {S}^{(i)}_t|\tilde {
\tau}_{(i)} > t\bigr],
\]
where $\tilde{\tau}_{(i)}$ is the stopping time of $\tilde{S}^{(i)}_t$.
Therefore, conditional on $\tau_{\alpha}, \tilde{\tau}_{(i)} > t$,
\[
\Ind\bigl[G_t^\alpha\bigl(\tilde{S}^{(i)}_t
\bigr) \leq\eta\bigr] \leq_{st} \Ind \bigl[G_t^\alpha
\bigl(S_t^\alpha\bigr) \leq\eta\bigr] \leq_{st} X,
\]
where $X$ is a Bernoulli variable with success probability $\eta$. It
follows that
\[
\sum_{i=r_t}^{n_t} \Ind\bigl[G_t^\alpha
\bigl(\tilde{S}^{(i)}_t\bigr) \leq\eta\bigr]
\leq_{st} B^+,
\]
where $B^+$ is a Binomial variable with success probability $\eta$ and
size $n_t -\break r_t +1$. Therefore, $T^+ \leq\tilde{T}^+\leq_{st}B^+$.
The bound for $T^-$ can be shown similarly.
\end{pf*}

\section{On the midpoint rule}

\begin{pf*}{Proof of Lemma~\ref{lemmidpoint}}
Let $t\in[0,1]$ and define $\Delta(t) = \sup\{\beta_2 - \beta_1\dvtx\break  \frac
{\beta_1 + \beta_2}{2} = t, \beta\in A\}$. This is well defined
because of (ii). The implication from left to right follows by the
definition of $\Delta$.\vadjust{\goodbreak}

Let $\beta\in C\dvtx \beta_2 -\beta_1 \leq\Delta(\frac{\beta_1 +
\beta_2}{2})$. Let $t = \frac{\beta_1 + \beta_2}{2}$. As $A$ is compact and
$D = \{\xi\in\mathbb{R}^2\dvtx \xi_1 + \xi_2 = 2 t\}$ is closed, $A
\cap
D$ is compact and thus $\{\beta_2 - \beta_1\dvtx \frac{\beta_1 + \beta_2}{2} = t, \beta\in A\}$ is compact also.

Hence, there exists a $\gamma\in A$ such that $(\gamma_2 + \gamma_1)/2
= t$ and $\gamma_2-\gamma_1 = \Delta(t)$. This implies that $\beta
\subseteq\gamma$ using (iii), implying that $\beta\in A$.
\end{pf*}
\end{appendix}

\begin{supplement}[id=suppA]
\stitle{Approximation of the optimal number of streams\\}
\slink[doi]{10.1214/12-AOS1076SUPP} 
\sdatatype{.pdf}
\sfilename{aos1076\_supp.pdf}
\sdescription{We describe a method that uses information from the
pilot sample to approximate
the expected effort of the algorithm as a function of the number $N$
of streams.
This method is used to
choose $N$. Its performance is illustrated in a
simulated experiment.}
\end{supplement}

%


\printaddresses


\begin{thebibliography}{11}

\bibitem{boos00}
%
\begin{barticle}[auto:STB|2013/01/23|16:20:06]
\bauthor{\bsnm{Boos},~\bfnm{D.}\binits{D.}} \AND
\bauthor{\bsnm{Zhang},~\bfnm{J.}\binits{J.}}
(\byear{2000}).
\btitle{Monte Carlo evaluation of resampling-based hypothesis tests}.
\bjournal{J. Amer. Statist. Assoc.}
\bvolume{95}
\bpages{486--492}.
\bptok{imsref}%
\end{barticle}
%
\endbibitem

\bibitem{clopper34}
%
\begin{barticle}[auto:STB|2013/01/23|16:20:06]
\bauthor{\bsnm{Clopper},~\bfnm{C.}\binits{C.}} \AND
\bauthor{\bsnm{Pearson},~\bfnm{E.}\binits{E.}}
(\byear{1934}).
\btitle{The use of confidence or fiducial limits illustrated in the
case of the
binomial}.
\bjournal{Biometrika}
\bvolume{26}
\bpages{404--413}.
\bptok{imsref}%
\end{barticle}
%
\endbibitem

\bibitem{davison1997bma}
%
\begin{bbook}[mr]
\bauthor{\bsnm{Davison},~\bfnm{A.~C.}\binits{A.~C.}} \AND
\bauthor{\bsnm{Hinkley},~\bfnm{D.~V.}\binits{D.~V.}}
(\byear{1997}).
\btitle{Bootstrap Methods and Their Application}.
\bseries{Cambridge Series in Statistical and Probabilistic Mathematics}
\bvolume{1}.
\bpublisher{Cambridge Univ. Press}, \blocation{Cambridge}.
\bid{mr={1478673}}
\bptok{imsref}%
\end{bbook}
%
\endbibitem

\bibitem{embrechts97}
%
\begin{bbook}[mr]
\bauthor{\bsnm{Embrechts},~\bfnm{Paul}\binits{P.}},
\bauthor{\bsnm{Kl{\"u}ppelberg},~\bfnm{Claudia}\binits{C.}} \AND
\bauthor{\bsnm{Mikosch},~\bfnm{Thomas}\binits{T.}}
(\byear{1997}).
\btitle{Modelling Extremal Events: For Insurance and Finance}.
\bseries{Applications of Mathematics (New York)}
\bvolume{33}.
\bpublisher{Springer}, \blocation{Berlin}.
\bid{mr={1458613}}
\bptok{imsref}%
\end{bbook}
%
\endbibitem

\bibitem{gandy09}
%
\begin{barticle}[mr]
\bauthor{\bsnm{Gandy},~\bfnm{Axel}\binits{A.}}
(\byear{2009}).
\btitle{Sequential implementation of {M}onte {C}arlo tests with uniformly
bounded resampling risk}.
\bjournal{J. Amer. Statist. Assoc.}
\bvolume{104}
\bpages{1504--1511}.
\bid{doi={10.1198/jasa.2009.tm08368}, issn={0162-1459}, mr={2750575}}
\bptok{imsref}%
\end{barticle}
%
\endbibitem

\bibitem{supplement}
%
\begin{bmisc}[auto:STB|2013/01/23|16:20:06]
\bauthor{\bsnm{Gandy},~\bfnm{A.}\binits{A.}} \AND
\bauthor{\bsnm{Rubin-Delanchy},~\bfnm{P.}\binits{P.}}
(\byear{2013}).
\bhowpublished{Supplement to ``An algorithm to compute the power of Monte Carlo tests with
guaranteed precision.''
DOI:\doiurl{10.1214/12-AOS1076SUPP}.}
\bptok{imsref}%
\end{bmisc}
%
\endbibitem

\bibitem{hoeffding63}
%
\begin{barticle}[mr]
\bauthor{\bsnm{Hoeffding},~\bfnm{Wassily}\binits{W.}}
(\byear{1963}).
\btitle{Probability inequalities for sums of bounded random variables}.
\bjournal{J.~Amer. Statist. Assoc.}
\bvolume{58}
\bpages{13--30}.
\bid{issn={0162-1459}, mr={0144363}}
\bptok{imsref}%
\end{barticle}
%
\endbibitem

\bibitem{keilson71}
%
\begin{barticle}[auto:STB|2013/01/23|16:20:06]
\bauthor{\bsnm{Keilson},~\bfnm{J.}\binits{J.}} \AND
\bauthor{\bsnm{Geber},~\bfnm{H.}\binits{H.}}
(\byear{1971}).
\btitle{Some results for discrete unimodality}.
\bjournal{J. Amer. Statist. Assoc.}
\bvolume{66}
\bpages{386--389}.
\bptok{imsref}%
\end{barticle}
%
\endbibitem

\bibitem{keilson82}
%
\begin{barticle}[mr]
\bauthor{\bsnm{Keilson},~\bfnm{Julian}\binits{J.}} \AND
\bauthor{\bsnm{Sumita},~\bfnm{Ushio}\binits{U.}}
(\byear{1982}).
\btitle{Uniform stochastic ordering and related inequalities}.
\bjournal{Canad. J. Statist.}
\bvolume{10}
\bpages{181--198}.
\bid{doi={10.2307/3556181}, issn={0319-5724}, mr={0691387}}
\bptok{imsref}%
\end{barticle}
%
\endbibitem

\bibitem{Oden91}
%
\begin{barticle}[auto:STB|2013/01/23|16:20:06]
\bauthor{\bsnm{Oden},~\bfnm{N.~L.}\binits{N.~L.}}
(\byear{1991}).
\btitle{Allocation of effort in Monte Carlo simulation for power of permutation
tests}.
\bjournal{J. Amer. Statist. Assoc.}
\bvolume{86}
\bpages{1074--1076}.
\bptok{imsref}%
\end{barticle}
%
\endbibitem

\end{thebibliography}
\end{document}